\documentclass[prl,twocolumn,superscriptaddress,preprintnumbers,showpacs,byrevtex,floatfix]{revtex4}
\usepackage{amsmath}
\usepackage{graphicx}
\usepackage{epsfig}

\begin{document}

\title{Ion trap transducers for quantum electromechanical oscillators}

\author{W. K. Hensinger}
\affiliation{FOCUS Center and Department of Physics, University of
Michigan, Ann Arbor, MI 48109-1120 USA}
\author{D. W. Utami}
\affiliation{Centre for Quantum Computer Technology and Department
of Physics, School of Physical Sciences, The University of
Queensland, QLD 4072 Australia}
\author{H.-S. Goan}
\affiliation{Department of Physics, National Taiwan University, Taipei
106 Taiwan}
\author{K. Schwab}
\affiliation{Laboratory for Physical Sciences, 8050 Greenmead
Drive, College Park MD 20740 USA}
\author{C. Monroe}
\affiliation{FOCUS Center and Department of Physics, University of
Michigan, Ann Arbor, MI 48109-1120 USA}
\author{G.~J.~Milburn}
\affiliation{Centre for Quantum Computer Technology and Department
of Physics, School of Physical Sciences, The University of
Queensland, QLD 4072 Australia}

\date{\today}

\begin{abstract}
An enduring challenge for contemporary physics is
to experimentally observe and control quantum behavior in
macroscopic systems. We show that a single trapped atomic ion
could be used to probe the quantum nature of a mesoscopic
mechanical oscillator precooled to 4K, and furthermore, to cool
the oscillator with high efficiency to its quantum ground state.
The proposed experiment could be performed using currently
available technology.
\end{abstract}


\maketitle

A quantum electromechanical system (QEMS) is a device where the
quantum nature of either the electronic or mechanical degrees of
freedom becomes important in the observable behavior
\cite{roukes,Blencowe:QEMS}. Potential applications of QEMS
include single spin detection \cite{Rugar}, single molecule mass
spectrometry \cite{Ekinci}, and readout for quantum information
devices \cite{Irish}. An excellent example of such a system is a
rf single electron-transistor integrated with a low loss, high
frequency nanomechanical resonator which demonstrated both
continuous position detection approximately a factor of five away
from the uncertainty principle limit, and direct observations of
the mechanical mode occupation factor as low as N$\approx$58
\cite{schwab2004}. However, this device suffers from both
practical and fundamental limitations; picowatt levels of
dissipation at the transistor is suspected to have blocked passive
cooling of the mechanics to lower temperatures and occupation
factors, as well as the fundamental sensitivity limitation due to
coupling to an intrinsically non QND (quantum nondemolition) variable such as
position.  Other methods to perform measurements on quantum
limited mechanical systems are clearly needed.  Furthermore,
coupling atomic systems to nano-electronic and nano-mechanical
devices appears to be a very promising and exciting new frontier,
where it is hoped one might combine the unmatched quantum
coherence and detection efficiencies of atomic physics, with the
sophisticated electronic and microstructures which are possible in
the condensed matter realm \cite{Madsen2004,Folmann}.

In this paper we investigate the use of a laser-cooled, trapped
atomic ion to both monitor and manipulate the number-state of a
QEMS oscillator, following earlier suggestions by Wineland and
others \cite{Heinzen,Nist_rev}.  As the temperature of an ion's
vibrational degree of freedom can be $<10^{-3}$ K, it is hoped
that such a transducer could generate far less thermal power than
low temperature single-electron devices. Furthermore, due to the
excellent optical readout achieved in ion trap systems via
fluorescence shelving \cite{Blatt88}, the ion trap transducer is
expected to posses sensitivity of QEMS energy at the level of a
single quanta, a detection which is very difficult with a simple
linear coupling to displacement.  Finally, we show that the
ion-QEMS system can be configured to provide a very effective
cooling mechanism for the mechanical oscillators \cite{Heinzen}
and estimate that it should be possible to cool a QEMS cantilever,
precooled only to 4 Kelvin, to its ground state with very high
efficiency.

\begin{figure}[h]
\centerline{\includegraphics[width=4cm]{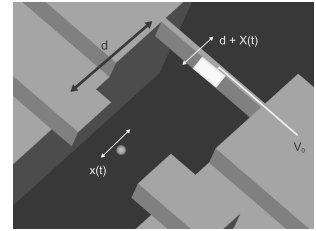}} \caption{Schematic
representation of a single trapped atomic ion coupled to an
nanoelectromecanical oscillator.} \label{fig-1}
\end{figure}

Here we describe a quantum model of a single trapped atomic ion
which is electrostatically coupled to a very small doubly clamped
cantilever, Fig. \ref{fig-1}. This coupling can be switched on and
off using an external bias voltage at an electrode on the
oscillator. The ion is held in a mesoscopic micro-fabricated
quadrupole Paul trap \cite{Madsen2004}, and laser cooled using
resolved sideband cooling \cite{Leibfried-RMP}. External lasers
are used to couple the internal electronic states of the ion to
its vibrational degree of freedom.
The bias gate on the oscillator carries a charge $Q=C_oV_o(t)$
where $C_o$ is the capacitance of the gate. We allow for the
possibility for the bias gate voltage $V_o(t)$ to be time
dependent so that it may be set to zero to turn off the
electrostatic coupling to the trapped ion.  The ion carries charge
$+e$. In the geometry of Fig. \ref{fig-1}, the interaction energy
between the ion and the oscillator is given by
$V_c=keV_oC_o/|d+\hat{X}(t)-\hat{x}(t)|$ where $d$ is the
equilibrium separation of the oscillator centre of mass position
and the ion. We set the equilibrium position of the ion at the
origin for simplicity. In that case $\hat{X}(t)$ and $\hat{x}(t)$
represent the small oscillations of the QEMS oscillator and the
ion around their equilibrium positions, respectively. We now
assume that the deviations from equilibrium are small compared to
the equilibrium separation $d$ and expand to second order in
$(\hat{X}(t)-\hat{x}(t))/d$. The interaction energy is then given
by $V_c=\left ( k e V_oC_o/d\right ) \left
[1-(\hat{X}(t)-\hat{x}(t))/d+(\hat{X}(t)-\hat{x}(t))^2/d^2\right
]$. The linear term may be absorbed into the definition
of the equilibrium positions. The quadratic term includes a coupling of the
oscillators and a renormalisation of the oscillation frequency for
both the ion and the oscillator. This frequency shift  is a small
perturbation of the bare frequencies and is neglected. We then see
that the interaction Hamiltonian coupling the ion to the
oscillator is $H_c= -\chi\hat{X}(t)\hat{x}(t)$ where
$\chi=2keV_oC_o/d^3$. A similar ion-oscillator system is described
in \cite{Tian2004}.

In ion trap technology a laser can be directed onto the ion to
induce transitions between the internal electronic states
\cite{Leibfried-RMP}. The electric field seen by the ion depends
on its motion in the trap, thus the external laser can couple the
internal electronic state to the vibrational degree of freedom. In
this way information on the vibrational motion can be transferred
to the electronic degree of freedom where it can be efficiently
measured using the technique of fluorescence shelving
\cite{Blatt88}. Such measurements are very nearly perfect
projective measurements onto one of two
electronic states, $\{|g\rangle,|e\rangle\}$. If the laser is
detuned to the first red (blue) motional sideband, the ion can
make a transition from the ground to the excited state absorbing a
laser photon and a single quanta of vibrational energy is
simultaneously added (deducted) in the process
\cite{Leibfried-RMP}. The probability of excitation depends on the
phonon distribution in the trap, so subsequent readout of the
electronic state then reveals information on the phonon number in
the trap.

The overall scheme for using the trapped ion as a transducer of
QEMS motion is as follows. The ion is cooled to the vibrational
ground state and prepared in some appropriate electronic state.
Next, a voltage is applied to the oscillator gate, coupling the
motion of the QEMS to the vibrational  state of the ion.
This transfers phonons from the oscillator to the ion. In the next
step an external laser couples the trap phonons onto the
electronic degree of freedom which is then readout using
fluorescence shelving.

We define dimensionless annihilation and creation operators for
the ionic vibration and the QEMS vibration using the ground state
standard deviation of position of the oscillator as a convenient
length scale in each case, $\hat{X} = \left
(\hbar/(2M\omega)\right )^{1/2}(a+a^\dagger)$ and $\hat{x} = \left
(\hbar/(2m\nu)\right )^{1/2}(b+b^\dagger)$ where  $\nu$ is the
vibrational frequency of the ion in the trap (also called the {\em
secular} frequency),  $\omega$ is the resonant frequency of the
QEMS oscillator, and $M,m$ are the masses of the QEMS and the ion
respectively. The total Hamiltonian for the three coupled systems
may be written as $H=\hbar\omega a^\dagger a+\hbar\nu b^\dagger
b-\hbar\kappa(a+a^\dagger)(b+b^\dagger)+\hbar\frac{\omega_A}{2}\sigma_z+H_{e}$
where $H_{e}$ describes the interaction between the external laser
and the electronic states of the ion (see below). The coupling
constant $\kappa$ is given by
$\kappa=(mM\nu\omega)^{-1/2}keV_oC_o/d^3$. The electronic
transition frequency is $\omega_A$ and $\sigma_z=|e\rangle\langle
e|-|g\rangle\langle g|$ is a Pauli matrix. The interaction between
the control laser and the electronic transition is given in the
dipole approximation by $H_e=2\hbar\Omega\sigma_x
\sin(k_l\hat{x}-\omega_l t)$ where $\omega_l$ and $k_l$ are the
frequency and wave vector of the laser, $\Omega$ is the effective
Rabi frequency for the transition and $\sigma_x=|e\rangle\langle
g|+|g\rangle\langle e|$. We will work in the Lamb-Dicke limit in
which the amplitude of the ion's motion in the direction of
radiation is much less than the wave length of the laser
$\sqrt{n_b+1}\eta=\sqrt{n_b+1}k_l \left (\hbar/(2m\nu)\right
)^{1/2} << 1$ in which case we may take
$H_e=2\hbar\eta\Omega\sigma_x (b+b^\dagger)\sin(\omega_l t)$. We
now move to the interaction picture and make the rotating wave
approximation for both the QEMS-ion coupling and the electronic
coupling, and assume that $\omega_l=\omega_A-\nu$ or
$\omega_l=\omega_A+\nu$ so that we are resolving the first red or
blue sideband. The total Hamiltonian is $H_I=\hbar\Delta a^\dagger
a-\hbar\kappa(a^\dagger b+ab^\dagger)+ H_{sb}$ with
$\Delta=\omega-\nu$ and $H_{sb}=\hbar
g(b\sigma_++b^\dagger\sigma_+^\dagger)$ for the red sideband and
$H_{sb}=\hbar g(b^\dagger\sigma_++b\sigma_+^\dagger)$ for the blue
sideband where $g=\eta\Omega$ and $\sigma_+=|e\rangle\langle g|$.
In our model both $\kappa$ and $g$ can be turned on and off. The
measurement protocol has two stages. In stage I, the oscillator
and cantilever are coupled for a time $\tau$ by setting
$\kappa\neq 0, g=0$. In stage II, $\kappa$ is turned off and the
electronic and vibrational degrees of freedom of the ion are
coupled by red and blue sideband excitation. At the end of this
stage, the ionic electronic state is measured.

Using the ion as an ultrasensitive sensor for the cantilever
motion is  experimentally realistic. Assuming a 19.7 MHz
cantilever, as reported recently by LaHaye et al.
\cite{schwab2004}, a radio-frequency microfabricated ion trap
\cite{Madsen2004} could be used to confine the ion at a distance
$\beta$ of $50 \mu m$ from the cantilever at a secular frequency
of 19.7 MHz. Applying a static voltage of 7.5V one would obtain a
coupling frequency $\Omega_d \approx 2 \pi\times52.5$ kHz for a
cadmium ion. Assuming the cantilever at a temperature at 4 K
(liquid Helium), the cantilever contains on the order of 1000
quanta of motion. Therefore one would only require an interaction
time of 5 $\mu s$ for the ion to undergo, on average, a single
phonon excitation. Anharmonicity in the cantilever motion is
included in the measured Q (too small to observe at 4K) and will
not measurably distort the cantilever lineshape for any reasonable
temperature \cite{Yurke1995}. An upper bound for the width of the
secular motion frequency due to the anharmonicity of the ion
motion at $\bar{n}_{b}=4000$ is found to be $\Delta\nu\approx \nu
(z/\beta)^2\sim$ where $z=(\hbar\bar{n}_{b}/(2m\nu))^{1/2}$is the
extent of the atomic wave function and $\beta$ is the dimension of
the ion trap. For the parameters considered here,
$\Delta\nu\ll\kappa$, so ion trap anharmonicities are not expected
to degrade the ion-cantilever coupling. The QEMS oscillator is
always coupled to a thermalising reservoir and before the coupling
to the ion is turned on we assume that the oscillator is in a
thermal equilibrium with mean vibrational quantum number
$\bar{n}_{a0}=(e^{\hbar\omega/k_BT}-1)^{-1}\approx
k_BT/(\hbar\omega)$.

When the coupling between the oscillator and the ion is turned on,
the ion becomes indirectly coupled to the phonon reservoir of the
QEMS oscillator.  This may be modelled using a quantum optics
master equation for the oscillator (if the frequency of the
oscillators is not too low) that describes the joint density
matrix operator $R(t)$ for the QEMS oscillator and the ionic
vibration motion during stage I,
\begin{eqnarray}
  \frac{dR}{dt}& = &-i \Delta[a^\dagger a, R] +i \kappa [a^\dagger
  b+ab^\dagger, R]
\nonumber \\
&& +\gamma_a \left(\bar{n}_{a0}+1\right) {\cal D}[ a] R+\gamma_a \bar{n}_{a0}
{\cal D}[ a^\dagger] R  \nonumber \\
&& + \mu_1 {\cal D}[b] R+ \mu_2 {\cal D} [b^\dagger] R,
\label{eq:masterEq1}
\end{eqnarray}
where $\cal{D}[O] \rho = O \rho O^{\dagger} -(O^{\dagger}O\rho+
\rho O^{\dagger}O)$/2 is defined for arbitrary operators $\cal{O}$
and $\rho$, and the damping rate $\gamma_a$ is related to the
quality factor $Q$ of the QEMS oscillator by the expression
$\gamma_a=\omega/Q$. Note that choosing the damping rates $\mu_1=
\gamma_b (\bar{n}_{b0}+1)$, $\mu_2=\gamma_b \bar{n}_{b0}$
corresponds to the dynamics for the ionic vibrational motion
induced by interacting with a thermal bath, as in the case for the
QEMS oscillator. However, if the main noise source coupled
directly to the vibrational modes of the trapped ion is the
fluctuating electric field generated in the trap electrodes
\cite{Turchette2000}, we can model it as a classical stochastic
electric field. In this case, we take $\mu_1=\mu_2=(1/\tau_1)$,
where $\tau_1$ is the characteristic heating time \cite{Budini02}.
Extrapolated from recent experiments and associated theory for
characteristic heating times \cite{Deslauriers2004,Turchette2000},
we obtain an expected approximate heating rate on the order of
0.06 quanta per ms for the above geometry for the cadmium ion.
Comparing this heating rate with the coupling constant $\kappa$
from our example above it is reasonable to neglect ion heating
during the cantilever-ion interaction. We are interested in the
mean phonon number $\bar{n}_{b}(\tau)=\langle b^\dagger (\tau)
b(\tau) \rangle$ of the thermal state of the ionic vibrational
modes at the end of stage one. Using Eq.~(\ref{eq:masterEq1}) the
equations of motion for quadratic moments form a closed set and do
not involve higher-order moments.

\begin{figure}[h]
\centerline{\includegraphics[width=8cm]{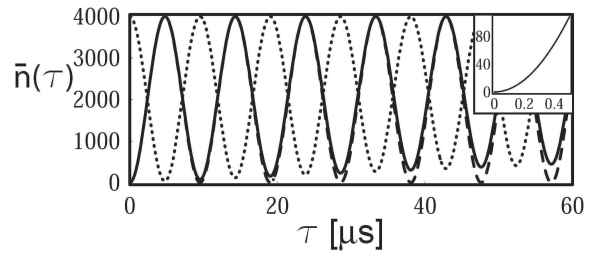}} \caption{The mean
phonon number of the ion $\bar{n}_{b}(\tau)$ as a function of time
$\tau$ obtained from Eq. (\ref{eq:ndamped}) is plotted (solid
line) and compared to the case where the coupling of cantilever
and thermal reservoir is neglected during the cantilever-ion
interaction (dashed line). The mean phonon number of the
cantilever $\bar{n}_a(\tau)$ is plotted as dotted line. The inset
shows quadratic short time behaviour of the ion.} \label{fig3}
\end{figure}

Neglecting heating due to a classical stochastic field during the
ion-cantilever interaction (as justified above) and assuming the
secular ion frequency equal to the resonance frequency of the
cantilever one can derive a simple analytical expression
($\Delta=\mu_1=\mu_2=0$):
\begin{eqnarray}
 \bar{n}_{b}(\tau)&=&\bar{n}_{a0}-\frac{(\bar{n}_{a0}-\bar{n}_{b0})}
{8\Omega_\gamma^2}  \, e^{-\gamma_a \tau/2} \nonumber \\
&&\times [(4 \Omega^2_\gamma-2\kappa^2+i\gamma_a\Omega_\gamma)
\, e^{-i2\Omega_\gamma \tau}+4\kappa^2 \nonumber\\
&& +(4 \Omega^2_\gamma -2\kappa^2-i \gamma_a\Omega_\gamma)\, e^{2
i\Omega_\gamma
 \tau}],
\label{eq:ndamped}
\end{eqnarray}
where $\Omega_\gamma=\sqrt{\kappa^2-(\gamma_a/4)^2}$. The initial
conditions used to obtain Eq.~(\ref{eq:ndamped}) assume that at
the start of stage one the QEMS oscillator and ionic vibrational
mode are in thermal states with mean phonon number $\bar{n}_{a0}$
and $\bar{n}_{b0}$, respectively. For $\gamma_a\ll\kappa$ one
obtains $
\bar{n}_{b}(\tau)=\bar{n}_{a0}-(\bar{n}_{a0}-\bar{n}_{b0})e^{-\gamma_a
\tau/2} cos^{2}(\kappa\tau)$. The mean oscillator phonon number
can also be found to be
$\bar{n}_{a}(\tau)=\bar{n}_{a0}-(\bar{n}_{a0}-\bar{n}_{b0}) \,
e^{-\gamma_a \tau/2} \, \kappa^2 \sin^2\left(\Omega_\gamma
\tau\right)/\Omega_\gamma^2$. In Fig. \ref{fig3}, we plot
$n_b(\tau)$ (obtained from Eq. (\ref{eq:ndamped})) using realistic
parameters corresponding to the mechanical cantilever of Ref.
\cite{schwab2004}. Assuming $\bar{n}_{a0}=4000$, $\bar{n}_{b0}=0$,
$\Delta=0$, $\kappa=2\pi\times 52.5$ kHz and assuming a Q factor
for the cantilever of 30000 we find that the effect of the
interaction of the cantilever with the thermal bath is
approximately negligible during the first quarter coupling period,
however, becomes significant on longer timescales. In the limit
$\gamma_a\tau<<1$ we find that
$\bar{n}_b(\tau)\approx\bar{n}_{a0}\sin^2(\kappa\tau)$.

From Fig. \ref{fig3} we see that eventually the ionic vibrational
degree of freedom comes into thermal equilibrium with the
cantilever. However due to the very strong coupling between the
two, the dynamics of this process is certainly not exponential.
The oscillations in Fig. \ref{fig3} are thus evidence of
non-exponential relaxation. The experiment proposed here provides
a convenient way to study the transition from exponential
relaxation to non-exponential short time behavior by varying the
coupling strength $\kappa$ with respect to the damping rate of the
cantilever, $\gamma_a$.

The readout process of stage II is similar to the experiments
described in Ref.~\cite{Leibfried-RMP}. In stage II we couple the
electronic state of the ion to its vibrational motion for a time
$T$ using the first red and blue sideband transitions. If we write
the probability for the atom to found in the excited state after
time $T$ as $P^R_e(T)$  and $P^B_e(T)$ for red and blue sideband
excitation respectively it can be shown
\cite{Monroe1995,Turchette2000} that the mean phonon number
$\bar{n}_{b}(\tau)$ is given by
$\bar{n}_{b}(\tau)/(1+\bar{n}_{b}(\tau))=P^R_e(T)/P^B_e(T)\equiv
R_e$. As this result is independent of the coupling time $T$,
noise due to heating during stage two can be kept small by
minimizing $T$. Indeed such experiments are routinely performed
\cite{Monroe1995,Roos1999,Deslauriers2004} and a phonon number
smaller than unity can be determined. As the parameters
$\kappa,\tau$ are assumed to be known this gives $\bar{n}_{a0}$
directly. In particular, for the short time region of Fig.
\ref{fig3} (and $\gamma_a\tau<<1$) we find that
$P^R_e(T)/P^B_e(T)\approx \bar{n}_{a0}\kappa^2 \tau^2$. Thus
measurement of the ratio of excitation probability on the first
red and blue sideband yields $\bar{n}_{a0}$ directly. Note that
$\bar{n}_{b}(\tau)$ should be on the order or smaller than
$n_{max}=20$ for reliable measurement
\cite{Nist_rev,Leibfried-RMP,Turchette2000}.

From Fig. \ref{fig3} we see that it is possible to almost
completely exchange the mean phonon number of the cantilever and
the ionic vibration. Since the ion begins in the vibrational
ground state, the result after one exchange time is to transfer
the thermal energy of the cantilever rapidly to the ionic
vibration, leaving the cantilever near its ground state.

A number of cooling procedures appear possible. One could simply
dump the resulting hot ion or one could implement two ion traps
that are located adjacent to the cantilever: one ion trap to be
used to cool the mechanics to the ground state, while the other
ion trap is used to perform the quantum measurement of the
cantilever. Alternately, an iterative cooling mechanism could be
accomplished by decoupling the resulting ion from the cold
cantilever (instead of dumping it) by detuning the ion trap
secular frequency and then laser cooling the ion. One could also
accomplish cooling of the cantilever and obtain a reasonable
temperature estimate using just a single ion provided
$\bar{n}_{a0} \pi \gamma_a/\kappa<n_{max}$ (meaning that the ion
carries less than $n_{max}$ phonons after one full coupling
period). In this scheme one waits for a full coupling period to
map out the minimum of $\bar{n}_{b}(\tau)$. Using this revival one
could deduce the optimum cooling interaction time (half the
coupling period) and the associated temperature of the cantilever.
Although the cold state is metastable for all of the schemes
above, with a lifetime determined by $\gamma_a$, we expect a
significantly lower resonator temperature compared to what has
been achieved with continuous refrigeration of the cantilever
bath, $\bar{n}_{a0}\sim50$ \cite{schwab2004}. Finally, one could
couple the ion continuously resulting in sympathetic cooling and
obtain a stable low temperature of the cantilever. This can achieve a
mean phonon number of the cantilever less than unity \cite{Tian2004}.

Approaching the mechanical ground state may be facilitated by
starting with a much higher frequency resonator (eg. a 1 Ghz
cantilever \cite{Huang}) and dilution refrigeration. While it may
be difficult to fabricate an ion trap with such high secular
frequency, it is possible to couple the cantilever to the
micromotion \cite{Nist_rev,Dehmelt67} in the ion trap that occurs
at $\Omega_{rf}+\nu$ and $\Omega_{rf}-\nu$ where $\Omega_{rf}$ is
the radio-frequency driving frequency of the ponderomotive ion
trapping potential. Micromotion could then be coupled via laser
excitation \cite{Sauter} into the electronic state of the ion.

This scheme is not a single shot readout of phonon number in the
oscillator, rather it enables a statistical inference of the mean
phonon number of the oscillator. Can it be used to measure a weak
classical force ? The sensitivity to changes in $\bar{n}_b(\tau)$
is best for values of $\bar{n}_b(\tau)$ less the unity. This can
be achieved by ensuring $\bar{n}_{a0}\kappa^2\tau^2\leq 1$. It
should then be possible to detect a change in the inferred mean
phonon number of the oscillator of the order of one quanta. If a
weak classical force is continuously applied to the cantilever
after it has been prepared close to the ground state (eg. via
sympathetic cooling via the ion), its equilibrium phonon
distribution will shift by an amount proportional to the square of
the ratio of applied force to the energy damping rate. Thus we
should be able to infer the size of a classical force so weak as
to shift the mean phonon number of the oscillator by one quanta.
This corresponds to a displacement sensitivity at the standard
quantum limit $\Delta x_{SQL}=(\hbar/2m\omega)^{1/2}$.

\acknowledgements{GJM acknowledges the support of the ARC
Federation Fellowship Grant. This work was supported by FOCUS seed
funding and the U.S. National Security Agency and Advanced Research and
Development Activity under Army Research Office contract
DAAD19-01-1-0667 and the National Science Foundation Information
Technology Research Program.}

\bibliographystyle{prsty}

\end{document}